\documentclass[conference]{IEEEtran}

\IEEEoverridecommandlockouts

\usepackage{cite}

\usepackage{multirow}

\ifCLASSINFOpdf
\usepackage[pdftex]{graphicx}
\else
\fi
\usepackage{subfigure}

\usepackage{amsmath}
\usepackage{algorithmic}
\usepackage{array}

\usepackage{fancyhdr}

\usepackage{float}
\usepackage{dblfloatfix}
\usepackage{enumitem}

\usepackage{url}
\begin{document}

\title{\vspace{-0.5em}\huge On QoS-Assured Degraded Provisioning in Service- Differentiated Multi-Layer Elastic Optical Networks\vspace{-0.3em}}

\author{
\IEEEauthorblockA{\textbf{
Zhizhen Zhong$^{1}$, Jipu Li$^{1}$, Nan Hua$^{1}$, Gustavo B. Figueiredo$^{3}$,
Yanhe Li$^{1}$, Xiaoping Zheng$^{1}$, Biswanath Mukherjee$^{2}$}}
\IEEEauthorblockA{$^1$Tsinghua National Laboratory for Information Science and Technology (TNList), \\Department of Electronic Engineering, Tsinghua University, Beijing 100084, China}
\IEEEauthorblockA{$^2$University of California, Davis, California 95616, USA \quad\quad $^3$Federal University of Bahia, Salvador, Brazil} 
\url{zhongzz14@mails.tsinghua.edu.cn},
\url{xpzheng@mail.tsinghua.edu.cn}
\thanks{This work is supported in parts by National 973 Program grant No. 2014CB340104/05, NSFC under grant No. 61201188, 61321004, Tsinghua Fudaoyuan Research Fund, and Networks Lab at UC Davis.}
\vspace {-1.5em}
}

\maketitle

\thispagestyle{fancy}
\chead{\small Accepted, IEEE Globecom 2016, Optical Networks and Systems (ONS) Symposium, for ArXiv only.}

\begin{abstract}
Degraded provisioning provides an effective solution to flexibly allocate resources in various dimensions to reduce blocking for differentiated demands when network congestion occurs. In this work, we investigate the novel problem of online degraded provisioning in service-differentiated multi-layer networks with optical elasticity. Quality of Service (QoS) is assured by service-holding-time prolongation and immediate access as soon as the service arrives without set-up delay. We decompose the problem into degraded routing and degraded resource allocation stages, and design polynomial-time algorithms with the enhanced multi-layer architecture to exploit network flexibility in temporal and spectral dimensions. Numerical results verify that we can achieve significant blocking reduction, especially for requests with higher priorities. They also indicate that degradation in optical layer can increase the network capacity, while degradation in electric layer provides flexible time-bandwidth exchange.
\end{abstract}

\IEEEpeerreviewmaketitle
\vspace {-0.5em}

\section{Introduction}
Network operators continuously upgrade their networks due to increasing demands for ubiquitous communications. As popularities of cloud computing, Internet of Things (IoT), and 5G mobile communications increase, network traffic is becoming heavy and extremely bursty. Thus, just enlarging the network capacity is not an economical choice, while neglecting it will strongly affect the QoS. Also, emerging applications, such as online gaming, data backups and virtual machine migrations, result in heterogeneous demands in telecom networks. Today's network users request more customized services, such as Virtual Private Network (VPN) and Video on Demand (VoD), and more differentiated demands with different prices \cite{he2005}. For some of the traffic which is delay-insensitive and can accept some compromise in bandwidth or other aspects, it can be preempted by more ``important" requests when the network becomes congested. Thus, to maintain cost-effectiveness and customers' loyalty, network operators can provide different grades of service besides sufficient bandwidth, instead of trying to support all the traffic without distinction \cite{muhammad2013}.

To address these problems, \emph{degraded provisioning} is proposed to provide a degraded level of service when network congestion occurs instead of no service at all. Generally, degraded provisioning has two approaches: 1) keep total amount of transferred traffic constant by time prolongation or modulation-level adjustment with immediate service access (QoS-assured), or 2) degrade requested bandwidth without time or modulation compensation, or no guarantee for immediate access (QoS-affected). We focus on QoS-assured degraded provisioning in this study. 

In a multi-layer network, QoS-assured degradation has different implementation methods in different layers. In electric layer, for a delay-insensitive and degradation-tolerant request, when network bandwidth is scarce, we degrade its transmission rate to enlarge the available bandwidth and extend its holding time based on the premise that total traffic amount is unchanged. Note that a traffic request cannot be degraded arbitrarily, and it is constrained by a given deadline \cite{fawaz2010}. In elastic optical layer, degradation refers to decreasing the number of occupied spectrum slots of a lightpath and raising the modulation level to guarantee the capacity. In OFDM-based elastic optical networks, modulation level can be dynamically reconfigured in DSP and DAC/ADC via software \cite{zhang2013}. But optical degradation has a constraint that higher-order modulation has shorter transmission reach \cite{bocoi2009}. 

Due to the flexibility enabled by degraded provisioning, there exist many studies on this topic in different kinds of optical networks. In WDM networks, Roy \emph{et al.} \cite{roy2008} studied degraded service using multipath routing in a QoS-affected way. Zhang \emph{et al.} \cite{zhang2010} studied reliable multipath provisioning, exploiting flexibility in bandwidth and delay. Andrei \emph{et al.}\cite{Andrei2010} proposed a deadline-driven method to flexibly provision services without immediate access. Savas \emph{et al.} \cite{sedef2014} introduced a dynamic scheme to reduce blocking by exploiting degraded-service tolerance in a QoS-affected way, and they also applied this method to increase network survivability \cite{sedef}. In Mixed-Line-Rate (MLR) networks, Vadrevu \emph{et al.} \cite{vadrevu2014} proposed a QoS-affected degradation scheme using multipath routing considering minimum-cost network design. But the ITU-T grid limit puts constraints on optical layer flexibility.

As a major development of optical technology, elastic optical networking enables more flexibility in optical modulation and spectrum allocation. Distance-adaptive spectrum allocation \cite{jinno2010} is a similar approach as optical degradation, but its limitations are that the modulation format of a lightpath is configured at one time and cannot be adjusted based on the fluctuation of traffic. Gkamas \emph{et al.} \cite{Gkamas2015} proposed a dynamic algorithm for joint multi-layer planning in IP-over-flexible-optical networks without considering dynamic adjustment of lightpath modulation. Recent progress in modulation format conversion \cite{huang2012} enables all-optical OOK to 16QAM adjustment, and its advantages in elastic optical networking were demonstrated in Yin \emph{et al.} \cite{Yin2014} with modulation-formats-convertible nodes. Thus, degraded provisioning can be extended to the elastic optical layer, and this important issue has not been fully understood, with no previous studies. 

We summarize our contributions as follows: 1) to the best of our knowledge, this is the first investigation on QoS-assured degraded provisioning problem in multi-layer networks with optical elasticity; 2) we propose an enhanced multi-layer network architecture and solve the complex dynamic degraded routing problem; and 3) we further propose novel dynamic heuristic algorithms in both layers to achieve degraded resource allocation. Results show that a significant increase of service acceptance can be achieved by multi-layer degradation. 

\thispagestyle{fancy}
\chead{\small Accepted, IEEE Globecom 2016, Optical Networks and Systems (ONS) Symposium, for ArXiv only.}
\section{Dynamic Degraded Provisioning Scheme}
We decompose the online degraded provisioning problem into two subroutines: 1) degraded routing, and 2) degraded resource allocation.

\subsection{Degraded Routing}
Degraded routing solves the subproblem of degraded-route computation when conventional routing cannot be performed due to resource shortage. Optical degraded routing acts similarly as electric degraded routing. The term \emph{request} here refers to the \emph{lightpath request} in optical layer, and the \emph{service request} in electric layer. There are two concerns: route hops (RH) and potential degraded requests (PDR). RH denotes the amount of resources occupied by the new request, while PDR denotes how many existing requests may be affected. 

We define a link in any layer of the multi-layer network as a tuple: $V_{ij,k}=(\mathbf{\Theta}, C)$, which is the $k^{th}$ link from $i$ to $j$. $\mathbf{\Theta}$ is a set that contains existing requests routed on this link, and $C$ is the available capacity of this link.  $r_{n}$ is a request, and $\mathcal{P}_{r_n}$ is a degraded route for $r_{n}$ in electric or optical layer. 
\vspace{-0.3em}
\begin{equation}
\small
\mathcal{P}_{r_n}=\{V_{ij,k}|r_n \in V_{ij,k}.\mathbf{\Theta}\}
\vspace{-0.4em}
\end{equation}

We introduce two metrics to evaluate the route $\mathcal{P}_{r_n}$. Note that the $|\cdot|$ operation returns the number of elements in a set.
\vspace{-0.5em}
\begin{equation}
\small
\mathcal{N}_{RH}=|\mathcal{P}_{r_{n}}|
\vspace{-0.3em}
\end{equation}
\begin{equation}
\small
\mathcal{N}_{PDR}=|\bigcup^{|\mathcal{P}_{r_{n}}|}_{c=1}\mathcal{P}_{r_{n}}[c].\mathbf{\Theta}|
\vspace{-0.5em}
\end{equation}

To calculate a route for minimizing RH, Dijkstra algorithm is applied. However, minimizing PDR is not that easy, because the minimizing-PDR problem aims to obtain a route that has the smallest PDR among all possible routes between a given source-destination $(s,d)$ pair. A straight-forward idea is to list all possible routes between a given $(s,d)$ pair and compare their PDR. But the complexity of this process is $O((N-2)!)$ ($N$ denotes the number of nodes). Here, we propose the enhanced multi-layer network architecture by introducing the auxiliary \emph{service layer}, which lies directly above the electric layer, and we solve the problem in polynomial time.

In the enhanced multi-layer architecture (Fig. 1), all nodes in the optical layer are mapped to upper two layers. There are two kinds of directional weighed links, i.e., request link and resource link. Request link weight is equal to a given large number times a binary that indicates whether there are existing requests between the node pair. Resource link weight is binary, which means if there is sufficient resource for this request. Note that, if the upper layer has isolated nodes (no connected edges), and the isolated node happens to be the source or destination of the request, then we replace the isolated node with the originating or terminating nodes of lightpaths (or requests) that bypass the isolated node. For example, in Fig. 1, $e^*$ is bypassed by request \#2 ($a^*$-$e^*$-$c^*$) on service layer. Thus, if we want to compute a degraded route on electric layer with minimized PDR for a new request ($a^*,e^*$), we should replace $e^*$ with $c^*$, and execute a shortest-path algorithm on service layer for the auxiliary request ($a^*,c^*$). Therefore, the \emph{minimizing-PDR problem} on one layer is transformed into a \emph{weighed shortest-path problem} on the upper layer. Finally, an actual route will be acquired based on the computed shortest path on upper layer.

\begin{figure}[t]
\centering
\vspace {-1em}
\includegraphics[width=2.5in]{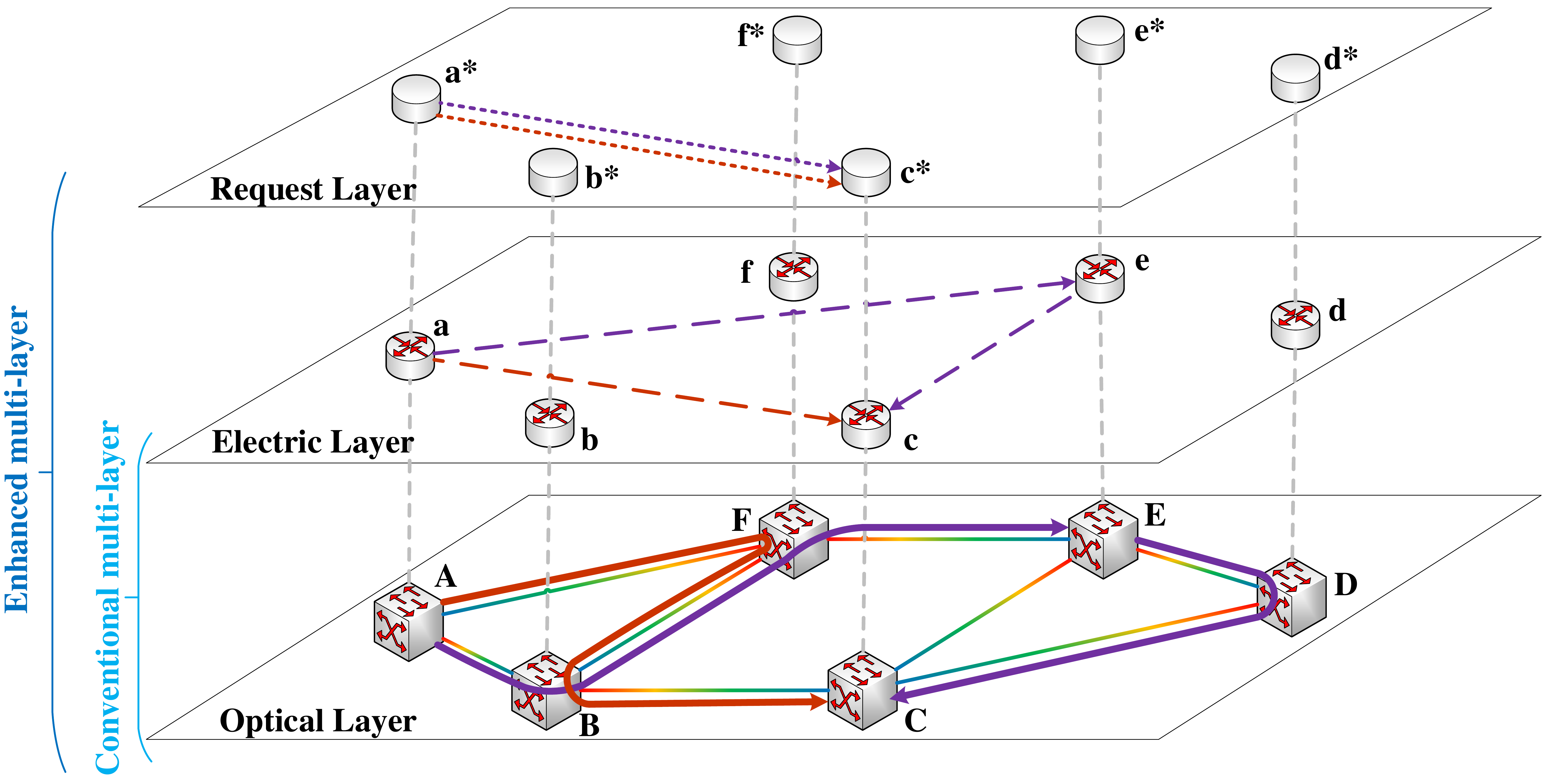}
\vspace {-0.5em} 
\caption{Illustration of enhanced multi-layer network architecture.}
\label{Algorithm graph}
\vspace {-1.7em}
\end{figure}

\begin{figure*}[t]
\centering
\vspace {-1.8em} 
\subfigure[Dynamic electric degradation.]{
\label{fig:subfig:a} %% label for first subfigure
\hspace{-3em}
\includegraphics[width=2in]{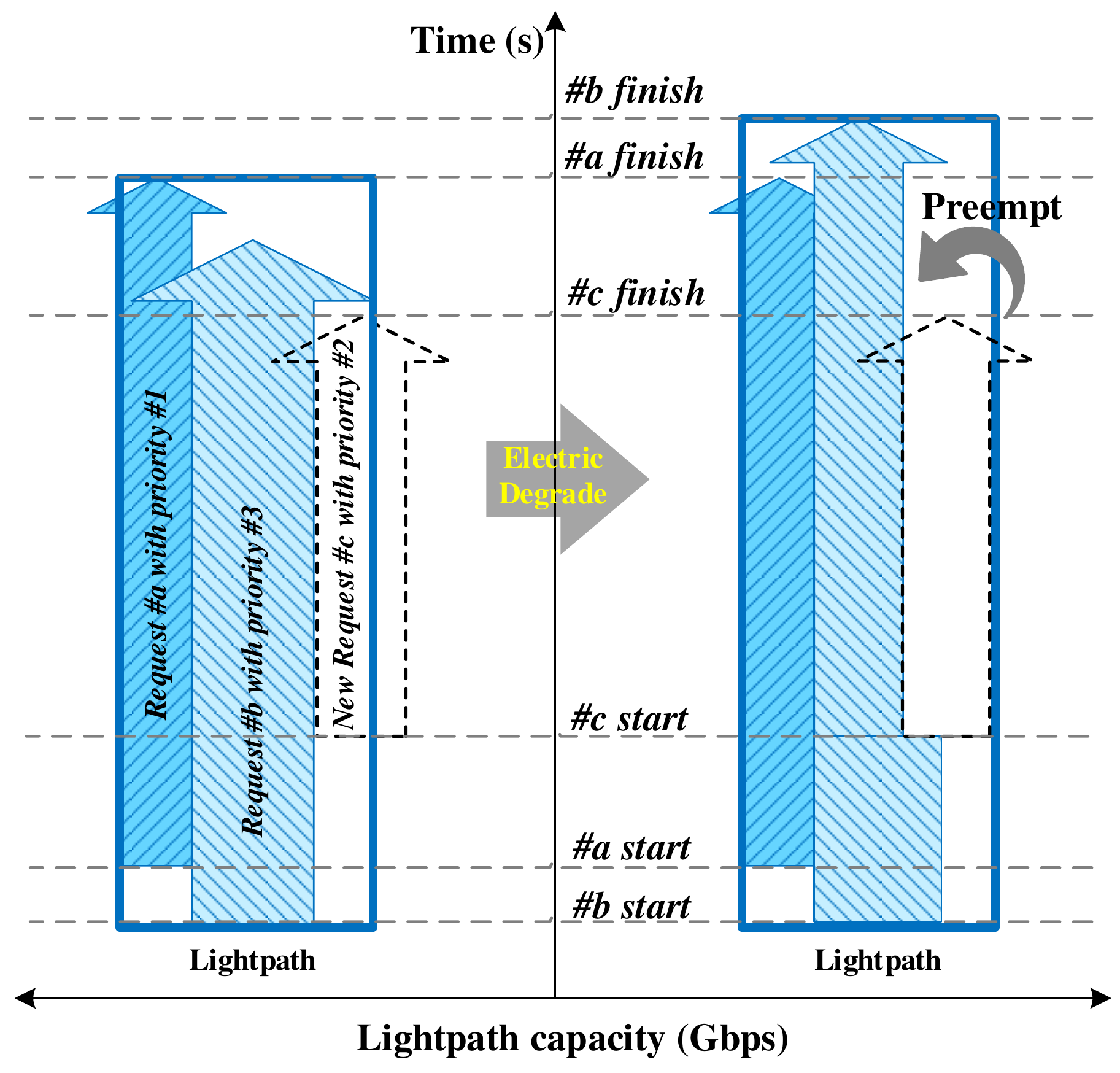}}
\hspace{0.5em}
\subfigure[Dynamic optical degradation: on a fiber.]{
\label{fig:subfig:a} %% label for first subfigure
\includegraphics[width=2in]{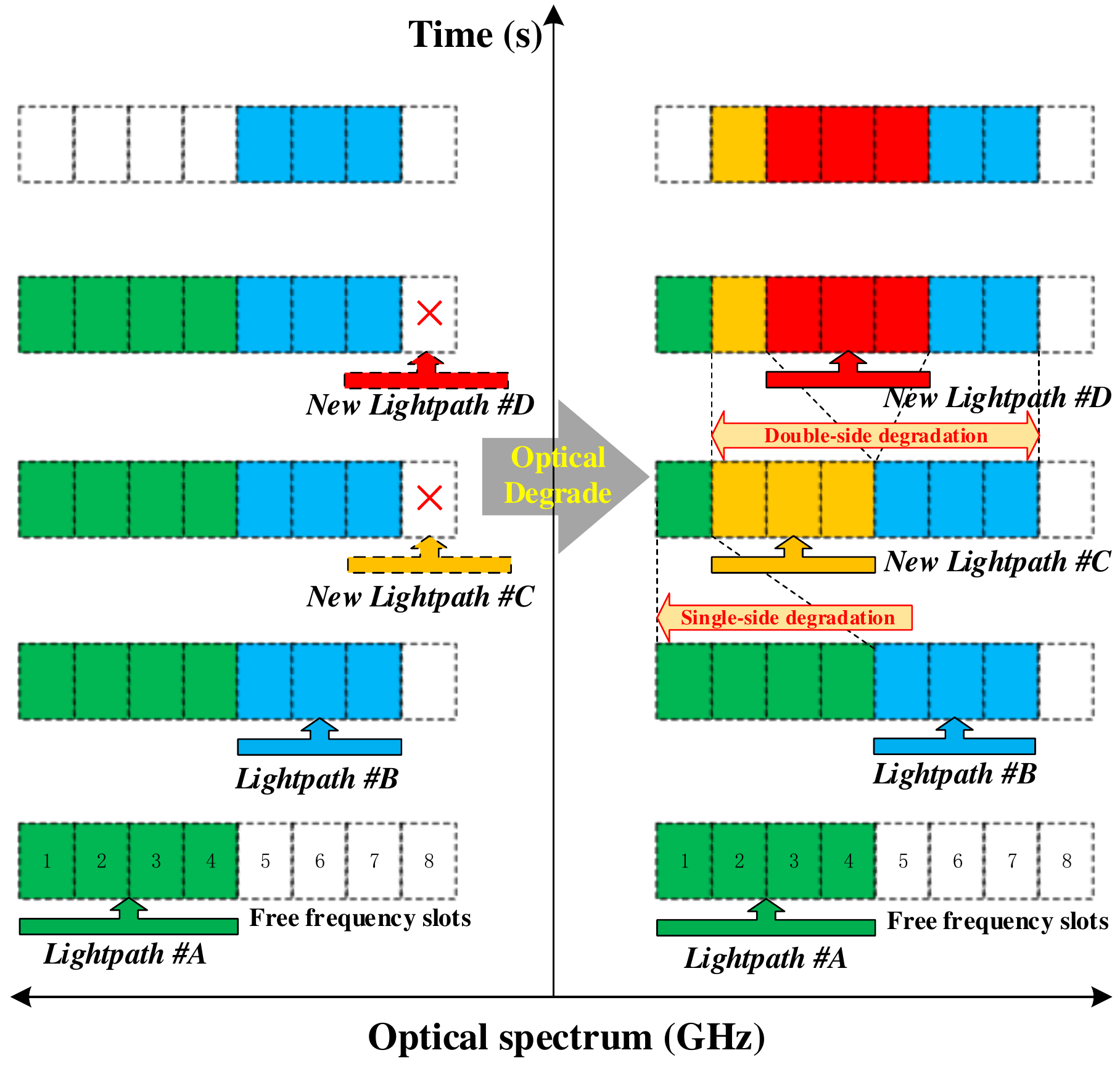}}
\hspace{0.5em}
\subfigure[Dynamic optical degradation: along a route.]{
\label{fig:subfig:a} %% label for first subfigure
\includegraphics[width=2.4in]{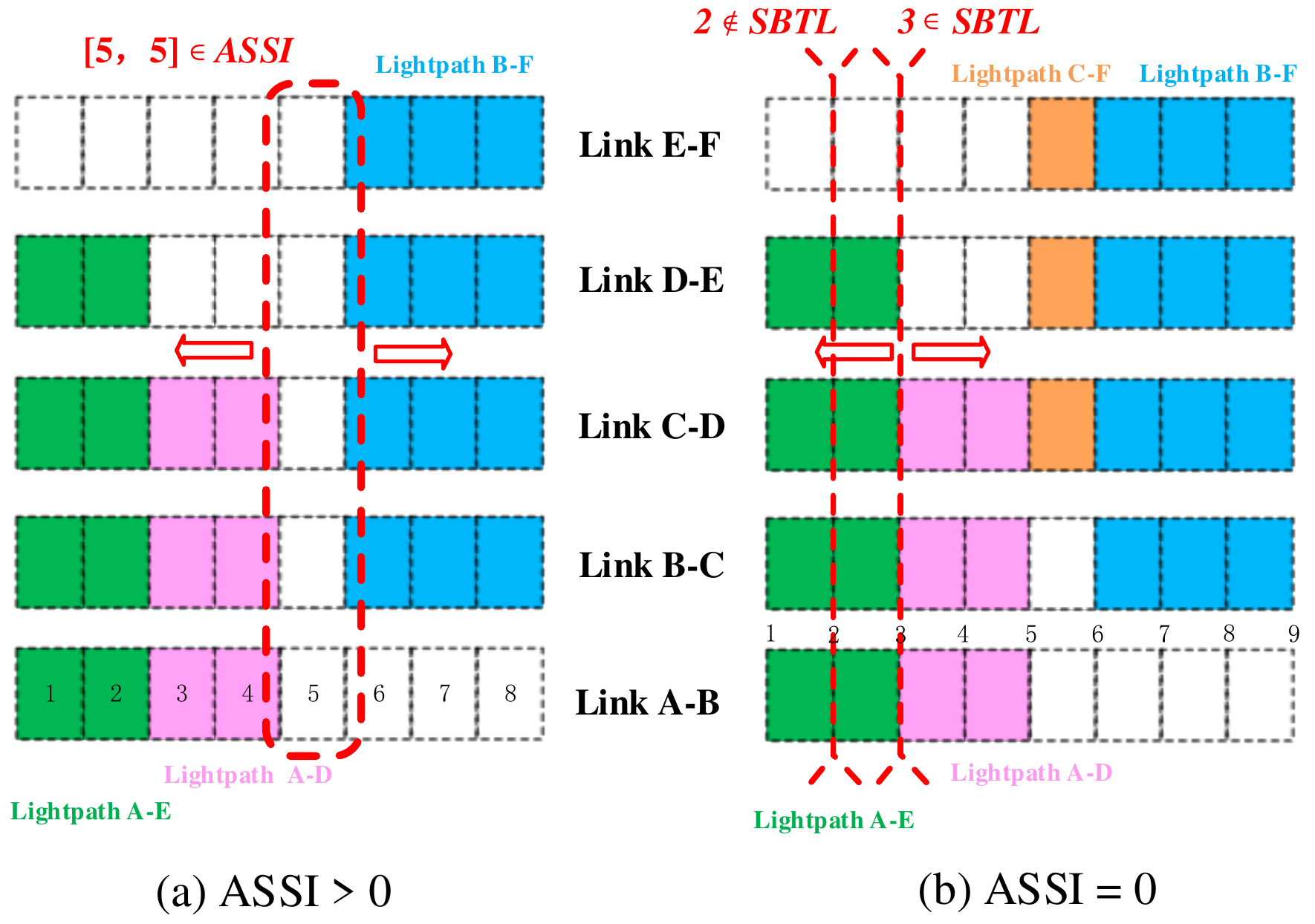}}
\hspace{-2.5em}
\vspace{-0.4em}
\caption{Illustration of multi-layer degradation principle in degraded resource allocation stage.}
\label{fig:subfig} %% label for entire figure
\vspace {-1.4em} 
\end{figure*}

\vspace {-1em}
\algsetup{ linenosize=\normalsize, linenodelimiter=. }
\noindent %
\rule [-0.3em] {\linewidth} {0.75pt} \\%
\noindent \textbf{Algorithm I}: Minimizing-PDR Algorithm
\\ \rule [0.8em]{\linewidth} {0.4pt} \\%
\vspace {-2.4em}
\begin{algorithmic}[1]
\footnotesize
\STATE Upper-layer topology$\{t_{i,j}\}=\{0\}$; current layer topology $\{t'_{i,j}\}$; $M$ is a large number; request ($s,d$);
\FOR {all requests $r$ \textbf{in} current layer}
\STATE $t_{r.source,r.destination}=M$; 
\ENDFOR
\FOR {all $i,j$} 
\STATE $t_{i,j}=t'_{i,j}$;
\ENDFOR
\IF {$\sum_{j \in N} t_{s,j}+\sum_{i\in N} t_{i,s}=0\ || \sum_{j \in N} t_{d,j}+\sum_{i\in N} t_{i,d}=0$} 
\STATE replace $s$ (or $d$) with the originating (or terminating) nodes of lightpaths (or request) that running bypass $s$ (or $d$);
\ENDIF
\STATE run shortest-path algorithm on upper-layer topology $\{t_{i,j}\}$, acquire $\mathcal{P}$;
\FOR {all links $V_{m,n}$ \textbf{in} $\mathcal{P}$} 
\STATE find a $V_{m,n}$ with the shortest hops (if $m$ or $n$ is the replaced node, use the isolated node to count hops) in lower layer; acquire its route;
\ENDFOR
\STATE combine all the acquired routes together, cancel loops and return it;
\end{algorithmic}
\vspace {-1em} %
\rule {\linewidth} {0.75pt} 
%\vspace {-1.2em}

Now, we introduce two policies of degraded routing:
\subsubsection{Minimize Route Hops (MinRH)}
We manage to minimize RH as a primary goal, and then minimize PDR. 

\subsubsection{Minimize Potential Degraded Requests (MinPDR)}
We try to minimize PDR first, then we minimize RH. 

\vspace{-0.3em}
\subsection{Degraded Resource Allocation}
When a degraded route $\mathcal{P}^e$ (electric) or $\mathcal{P}^o$ (optical) is acquired, we need to decide which request or requests to degrade, and how much to degrade them. In the multi-layer network, degraded resource allocation refers to different operations in different layers, which should be further studied separately.

\subsubsection{Electric Degraded Bandwidth Allocation (ED-BA)}
We propose the ED-BA algorithm based on a degraded route $\mathcal{P}^e$. A traffic service request on electric layer is defined as: $r_{n}=(s, d, bw, t, \tau, \eta, \rho)$, which mean source, destination, bandwidth, arrival time point, holding time, prolongation deadline, and priority, respectively. We define a function $AS(S, k)$, which sorts elements in set $S$ in ascending order of $k$.

\vspace {-1em}
\algsetup{ linenosize=\normalsize, linenodelimiter=. }
\noindent %
\rule [-0.3em] {\linewidth} {0.75pt} \\%
\noindent \textbf{Algorithm II}: ED-BA Algorithm
\\ \rule [0.8em]{\linewidth} {0.4pt} \\%
\vspace {-2.4em}
\begin{algorithmic}[1]
\footnotesize
\STATE Current time $t_c$, arriving request $r_0$, $flag$ = 1; 
\FOR {all links $V_n$ \textbf{in} $\mathcal{P}^e$}
\IF{$V_n.C < r_0.bw$}
\STATE $PDL(V_n)\leftarrow \{r_0\}$; \quad \emph{/*potential degraded links*/}
\FOR {all requests $r$ \textbf{in} $V_n.\mathbf{\Theta}$}
\IF {$r.\rho \leq r_0.\rho$}
\STATE $PDL(V_n).pushback(r)$;
\ELSE 
\STATE continue;
\ENDIF
\ENDFOR
\IF {$\sum_{u \in PDR(V_n)} u.bw\times\frac{u.\eta}{(u.\tau+u.\eta)} \geq r_0.bw$}
\STATE $ac\_bw\leftarrow$ 0; \ \emph{/*accumulate available bandwidth*/}
\FOR {all requests $x$ \textbf{in} $AS(PDL(V_n), priority)$}
\STATE degrade $x.bw$ to its maximum extent $x.bw'$, s.t. $\frac{(x.bw\times x.\tau-(t_c-x.t)\times x.bw)}{x.bw'}+t_c-x.t \leq x.\eta+x.\tau$;
\STATE $ac\_bw= ac\_bw+x.bw-x.bw'$;
\IF {$ac\_bw > r_0.bw$}
\STATE request $r_0$ routed successfully on $V_n$; break;
\ENDIF
\ENDFOR
\ELSE
\STATE request $r_0$ blocked; $flag$ = 0; break;
\ENDIF
\ELSE
\STATE continue;
\ENDIF
\ENDFOR
\IF {$flag$ == 1}
\STATE request $r_0$ is routed successfully;
\ENDIF
\end{algorithmic}
\vspace {-1em} %
\rule {\linewidth} {0.75pt} 
%\vspace {-1em}

 Fig. 2(a) shows the basic principle of the ED-BA algorithm, that requests with higher priorities can ``preempt" those  with no higher priorities. Here, the term ``preempt" means that some existing requests are degraded in transmission rate due to their relatively low priorities. Meanwhile, we manage to degrade the minimal number of requests to provide just-enough bandwidth. 
\thispagestyle{fancy}
\chead{\small Accepted, IEEE Globecom 2016, Optical Networks and Systems (ONS) Symposium, for ArXiv only.}
\subsubsection{Optical Degraded Modulation and Spectrum Allocation (OD-MSA)}
In an elastic optical network, optical degradation refers to the reduction of occupied-spectrum-slot number of a lightpath, and raise lightpath's modulation level to ensure lightpath capacity under the modulation-distance constraint.

Figs. 2(b) and 2(c) show how optical degradation works. The number of spectrum slots in a fiber is $B$. $S_f$ is a binary bitmask that contains $B$ bits to record the availability of each spectrum slot in fiber $f$, e.g., $S_f[p]=1$ means the $p^{th}$ slot is utilized. A lightpath is defined as a tuple: $L=(f, \xi_l, \xi_r, \eta, \delta)$, which denotes the fiber that the lightpath is routed through, left and right indices of occupied spectrum slots, modulation level, and lightpath distance, respectively. Note that $(\xi_r-\xi_l+1)\cdot log_2\eta$ should be constant when performing optical degradation. A lightpath request is defined as a tuple: $l_n=\{i, j, \theta\}$, which denotes source, destination, and requested bandwidth in spectrum slots. We define a function $Q(a)$ to get the transmission reach of modulation level $a$. 

We define Available Spectrum Slots Intersection (ASSI), which is a set of slots that are available all along the optical degraded route $\mathcal{P}^o$, to evaluate the available resources for the new lightpath before optical degradation. Note that the operator $\lor$ represents the logical OR operation, and $p \in [1,B]$.
\vspace {-0.5em}
\begin{equation}
\small
ASSI=\{S_f[p] | \sum_{f =1}^{|\mathcal{P}^o|-1}S_f[p] \lor S_{f+1}[p]=0\}
\vspace {-0.5em}
\end{equation}

We define Slot Border Through Lightpaths (SBTL) to evaluate whether a slot border locates inside the occupied spectrum of a lightpath. $w$ denotes index of spectrum slot borders, and there are $B+1$ borders, thus $w \in [1, B+1]$. We define a decision function $D(x)$ which is equal to 0 if $x$ is positive, or returns 1 if $x$ is negative.
\vspace {-0.5em}
\begin{equation}
\small
SBTL=\{w|\sum_{L:L.f\in\mathcal{P}^o}D((w-\frac{1}{2}-L.\xi_l)(w-\frac{1}{2}-L.\xi_r))=0\}
\vspace {-0.5em}
\end{equation}

When a degradation location (available slots or a slot border) is found in the spectrum by ASSI or SBTL, the potential degraded lightpaths are those on both sides of the location. We first try to degrade the left one ($L_1$), which is called \emph{single-side degradation}. And if only $L_1$ degradation cannot provide enough slots, we continue to degrade the right one ($L_2$), which is called \emph{double-side degradation}. Note that, when there are multiple possibilities (choosing an element in ASSI or SBTL), a First-Fit policy is applied to choose the one with smaller index to reduce spectrum fragmentation. 

On new lightpath establishment, we use the threshold-based grooming approach, which has been demonstrated to achieve lower blocking probability than the fixed-grid IP-over-WDM and elastic-spectrum non-grooming approaches \cite{wan2012}. 

\vspace {-1em}
\algsetup{ linenosize=\normalsize, linenodelimiter=. }
\noindent %
\rule [-0.3em] {\linewidth} {0.75pt} \\%
\noindent \textbf{Algorithm III}: OD-MSA Algorithm
\\ \rule [0.8em]{\linewidth} {0.4pt} \\%
\vspace {-2.4em}
\begin{algorithmic}[1]
\footnotesize
\STATE Arriving lightpath request $l_0$;
\FOR {all fibers $f_n$ \textbf{in} $\mathcal{P}^o$}
\STATE scan the spectrum to acquire $ASSI$;
\IF{$|ASSI|>0$} 
\STATE choose consecutive available slots $[l,r]$ with the largest $r-l$ value in $ASSI$;
\STATE scan the spectrum to acquire $L_1$ and $L_2$ ($L_1.\xi_r==l-1\ \&\&\ L_2.\xi_l==r+1$);
\IF {$Q(log_2a+1)< L_1.\delta \leq Q(log_2a)$ and $Q(log_2b+1)< L_2.\delta \leq Q(log_2b)$}
\STATE $L_1.\eta =a$; \quad\quad \emph{/*degrade $L_1$ to modulation level $a$*/}
\IF {$(L_1.\xi_r-L_1.\xi_l+1)(1-log_2(\eta_0-a))+(l-r+1)\geq l_0.\theta$}
\STATE setup a new lightpath with $l_0.spt$ slots, starting from $L_1.\xi_l+(L_1.\xi_r-L_1.\xi_l+1)log_2(\eta_0-a)$;
\ELSIF{$(L_1.\xi_r-L_1.\xi_l+1)(1-log_2(\eta_0-a))+(l-r+1)+(L_2.\xi_r-L_2.\xi_l+1)(1-log_2(\eta_0-b))\geq l_0.\theta$}
\STATE $L_2.\eta =b$;  \quad\emph{/*degrade $L_2$ to modulation level $b$*/}
\STATE setup a new lightpath with $l_0.\theta$ slots, starting from $L_1.lft+(L_1.\xi_r-L_1.\xi_l+1)log_2(\eta_0-a)$;
\ELSE
\STATE request $l_0$ blocked; break;
\ENDIF
\ENDIF
\ELSIF {$|SBTL|>0$}
\STATE choose smallest $w$ in $SBTL$, and perform sentence 7 to 17 (here, let $l=r+1=w$);
\ELSE
\STATE request $l_0$ blocked; break;
\ENDIF
\ENDFOR
\end{algorithmic}
\vspace {-1em} %
\rule {\linewidth} {0.75pt} 

\vspace {-0.6em}
\subsection{Complexity Analysis}
In degraded routing stage, minimizing-RH problem can be solved with Dijkstra algorithm with $O(N^2)$ complexity in a $N$-node topology. But the minimizing-PDR problem may have a complexity of $O(N^3)$ if both source and destination nodes of the request are isolated nodes\footnote{Here, the isolated nodes in request should be replaced by other nodes of the upper layer in path computation, and the problem becomes an all-pairs shortest-path one, which should be solved via Floyd algorithm ($O(N^3)$).}.

In degraded resource allocation stage, the worst case is that the degraded route goes through  every node of the topology, and the complexity is $O(N)$. In ED-BA algorithm, we suppose that the maximum number of existing requests on each link is $R$ (related to traffic load), and the time complexity is $O(NR)$. In OD-MSA algorithm, the complexity\footnote{Here,  $B$ is the number of spectrum slots, which is a constant parameter.} is $O(NB^2)$.

Hence, the complexity of the proposed dynamic degraded provisioning scheme is $O(N^3+NR)$, and is suitable for online decision making in dynamic traffic accommodation.
\thispagestyle{fancy}
\chead{\small Accepted, IEEE Globecom 2016, Optical Networks and Systems (ONS) Symposium, for ArXiv only.}
\section{Illustrative Numerical Evaluations}

\begin{figure}[!t]
\centering
\vspace {-1em}
\includegraphics[scale=0.22]{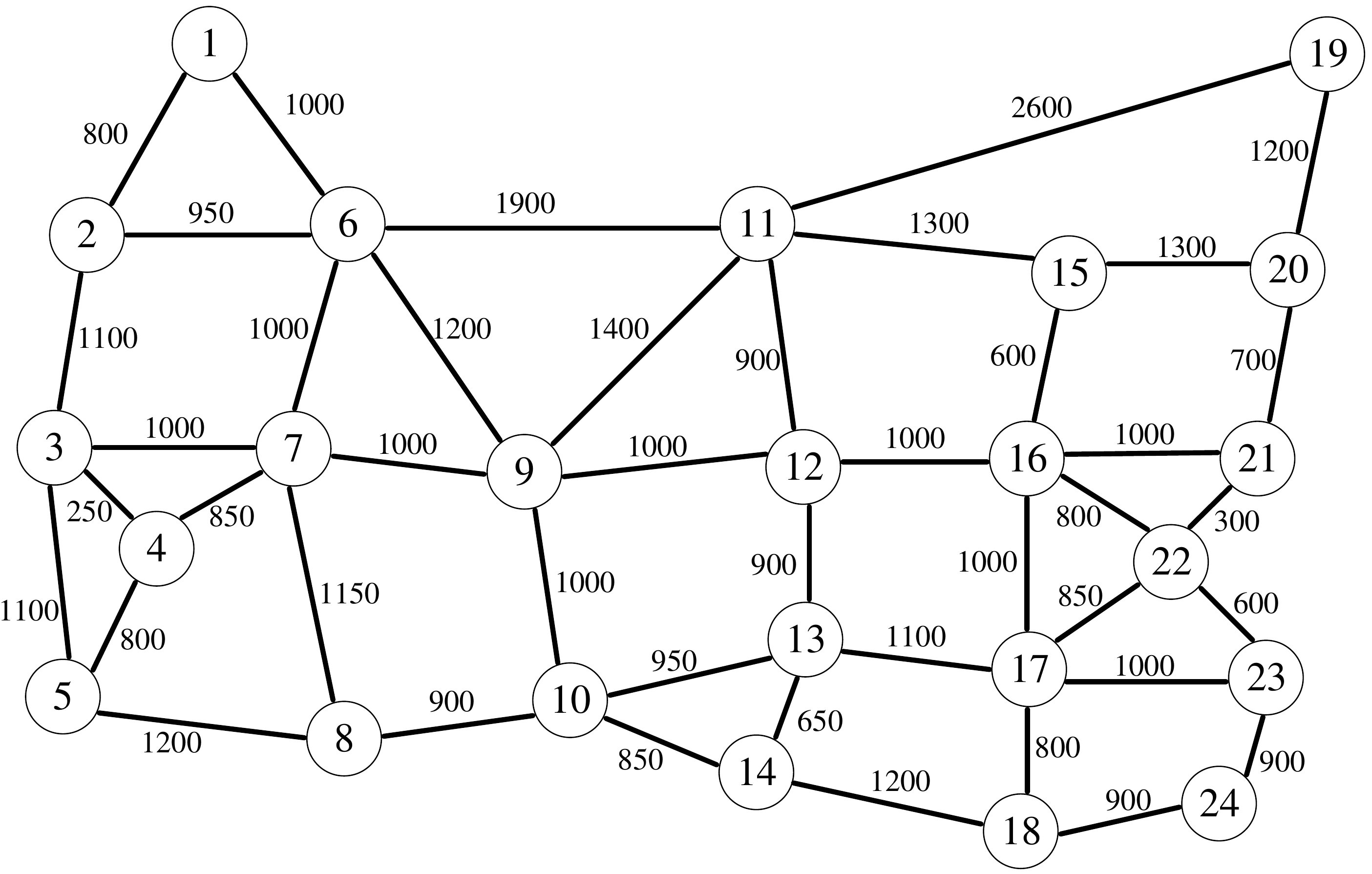}
\vspace {-1em}
\caption{USNet topology with fiber length in kilometers marked on links.}
\label{Algorithm graph}
\vspace {-1.5em}
\end{figure}

\subsection{Experimental Setup}
Table I summarizes the parameters of different modulation formats based on \cite{bocoi2009} \cite{zhu2013}. We consider the USNet topology (Fig. 3) for dynamic performance simulation. All fibers are unidirectional with 300 spectrum slots, and spectrum width of each slot is 12.5 GHz. Traffic requests are generated between all node pairs, and characterized by Poisson arrivals with negative exponential holding times. Granularities of requests are distributed independently and uniformly from 5 Gbps to 150 Gbps. The maximum acceptable ratio for transmission-rate degradation is uniformly distributed in [25\%, 100\%] \cite{sedef}. There are 5 priorities with equal amount each. The lightpath establishment threshold for grooming is chosen as 150 Gbps, which is equal to the largest request bandwidth and performs the best \cite{wan2012}. An event-driven dynamic simulator has been developed to verify the effectiveness of the heuristic algorithms. Six degradation policies, i.e. OE-MinPDR, O-MinPDR, E-MinPDR, OE-MinRH, O-MinRH, E-MinRH (OE: both-layer degradation, O: optical degradation only, E: electric degradation only) are studied.

\begin{figure*}[!t]
\centering
\vspace{-1em}
\hspace{-2em}
\subfigure[Requests with all priorities.]{
\label{fig:subfig:a} %% label for first subfigure
\includegraphics[width=2.2in]{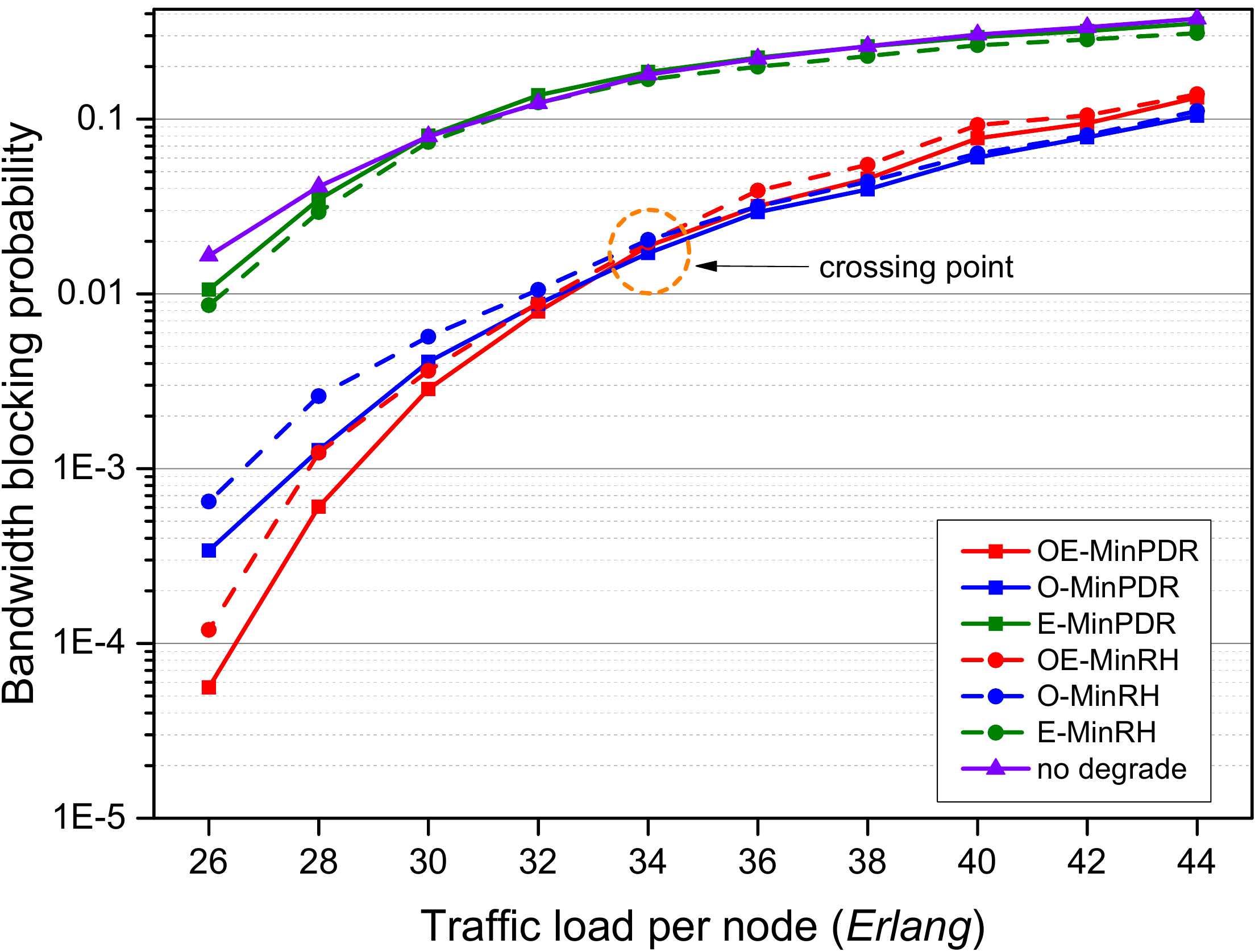}}
\hspace{0.2em}
\subfigure[Requests with highest priority (\#5).]{
\label{fig:subfig:a} %% label for first subfigure
\includegraphics[width=2.2in]{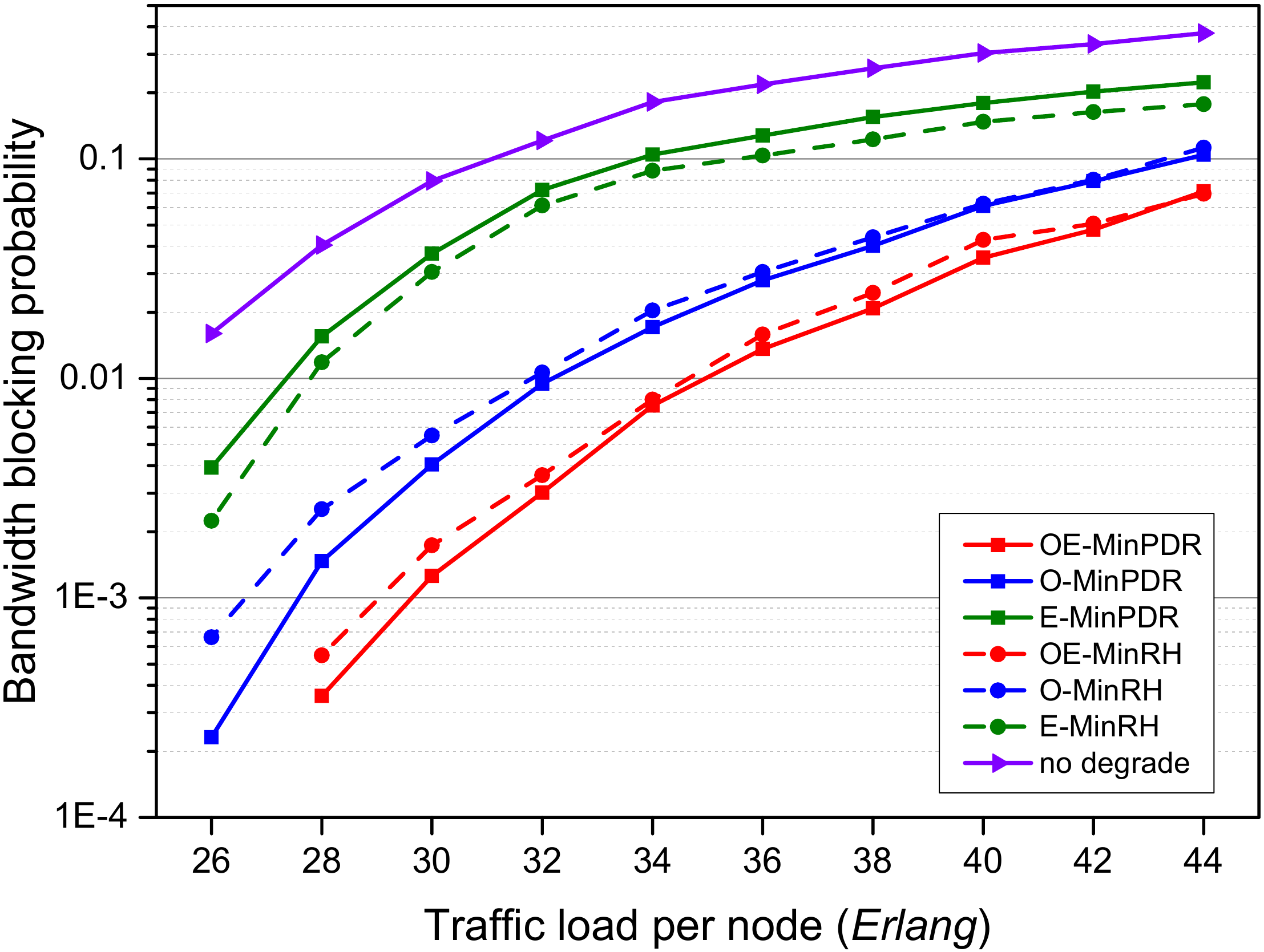}}
\hspace{0.2em}
\subfigure[Requests with lowest priority (\#1).]{
\label{fig:subfig:a} %% label for first subfigure
\includegraphics[width=2.2in]{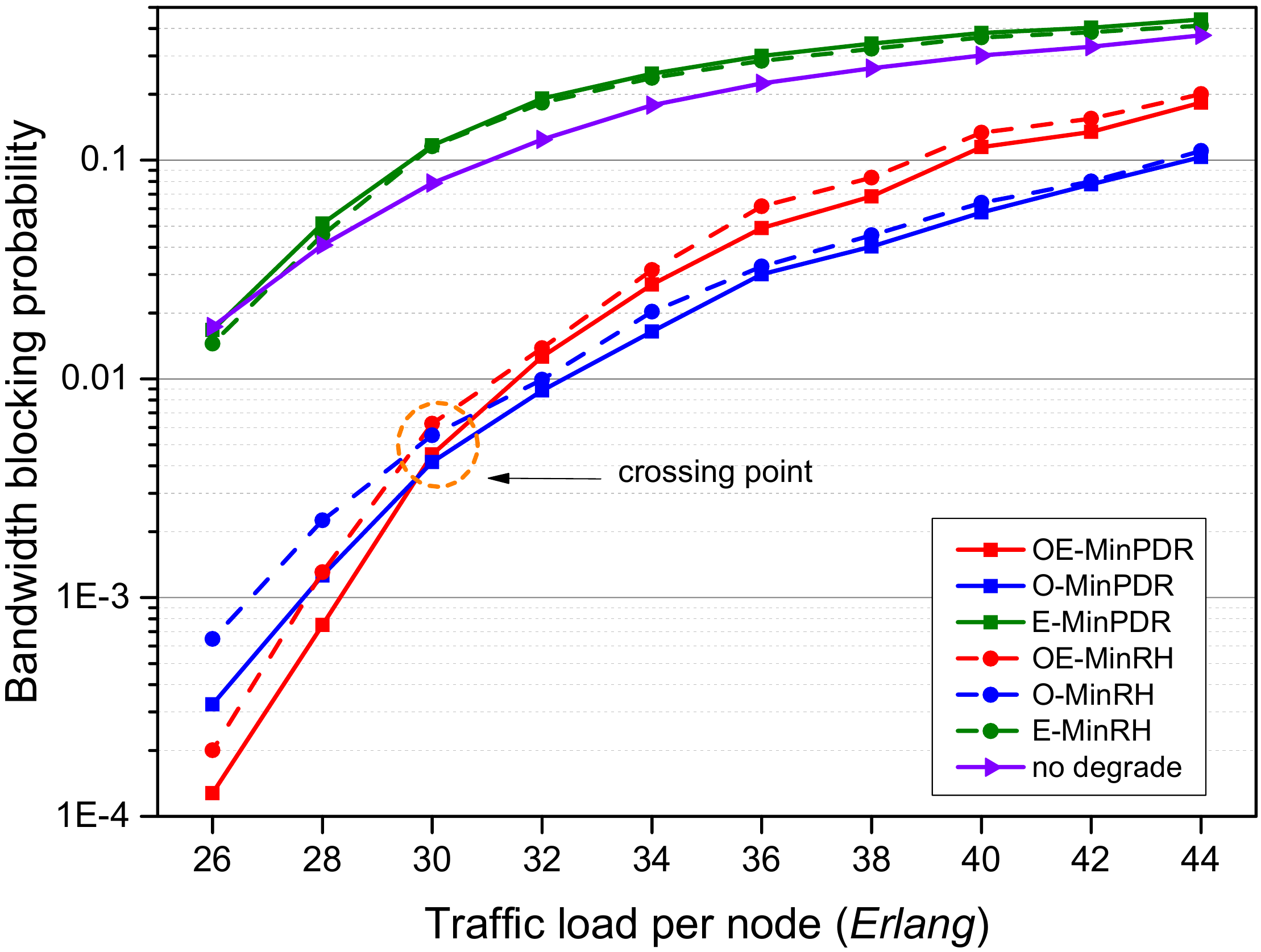}}
\hspace{-2em}
\vspace{-0.5em}
\caption{Bandwidth blocking probability vs. traffic load: different degradation policies.}
\label{fig:subfig} %% label for entire figure
\vspace {-1em} 
\end{figure*}

\begin{figure*}[!t]
\centering
\hspace{-2.5em}
\subfigure[Instantaneous network throughput (MinPDR).]{
\label{fig:subfig:a} %% label for first subfigure
\includegraphics[width=2.3in]{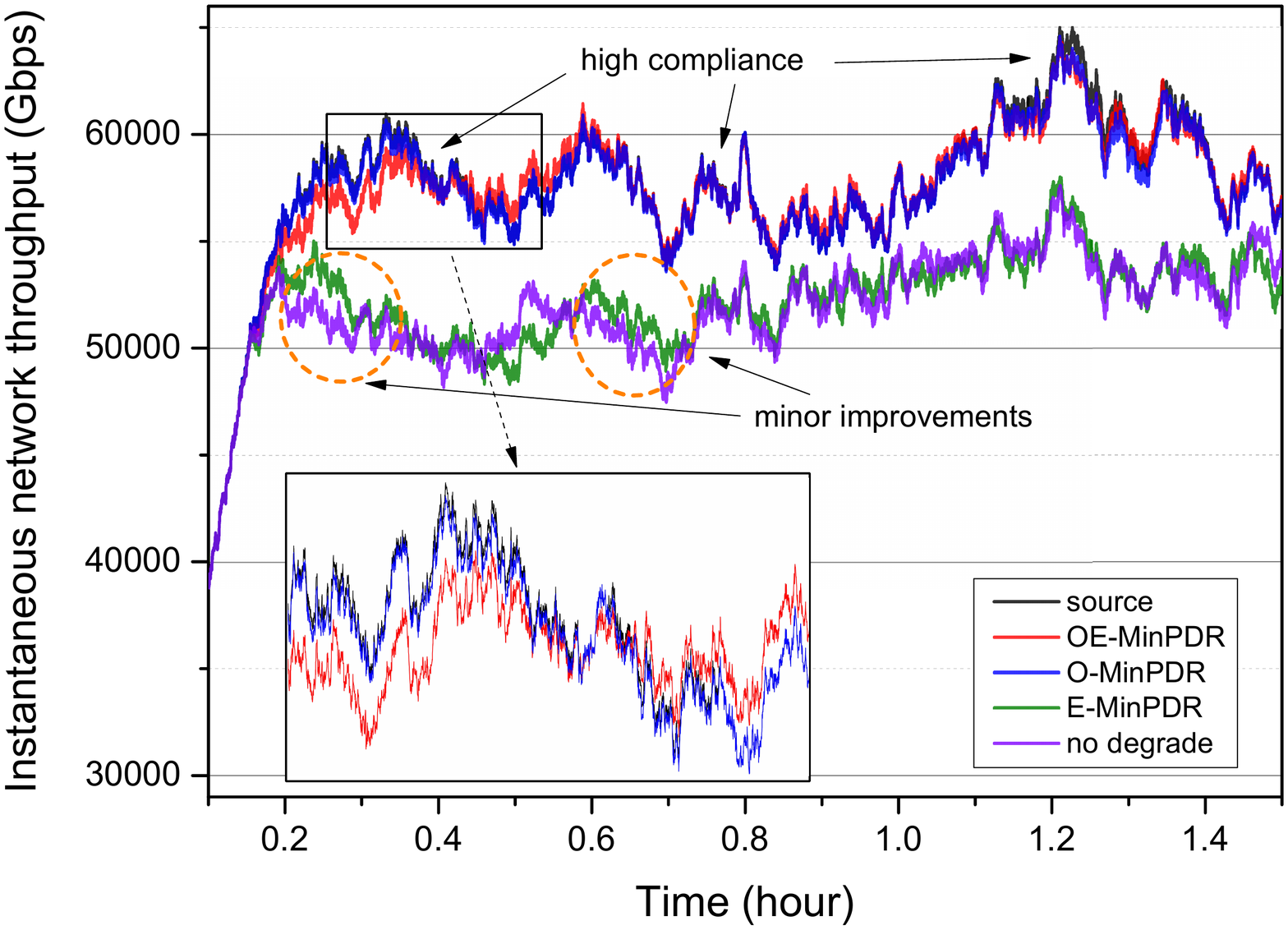}}
\hspace{0.15em}
\subfigure[Instantaneous network throughput (MinRH).]{
\label{fig:subfig:a} %% label for first subfigure
\includegraphics[width=2.3in]{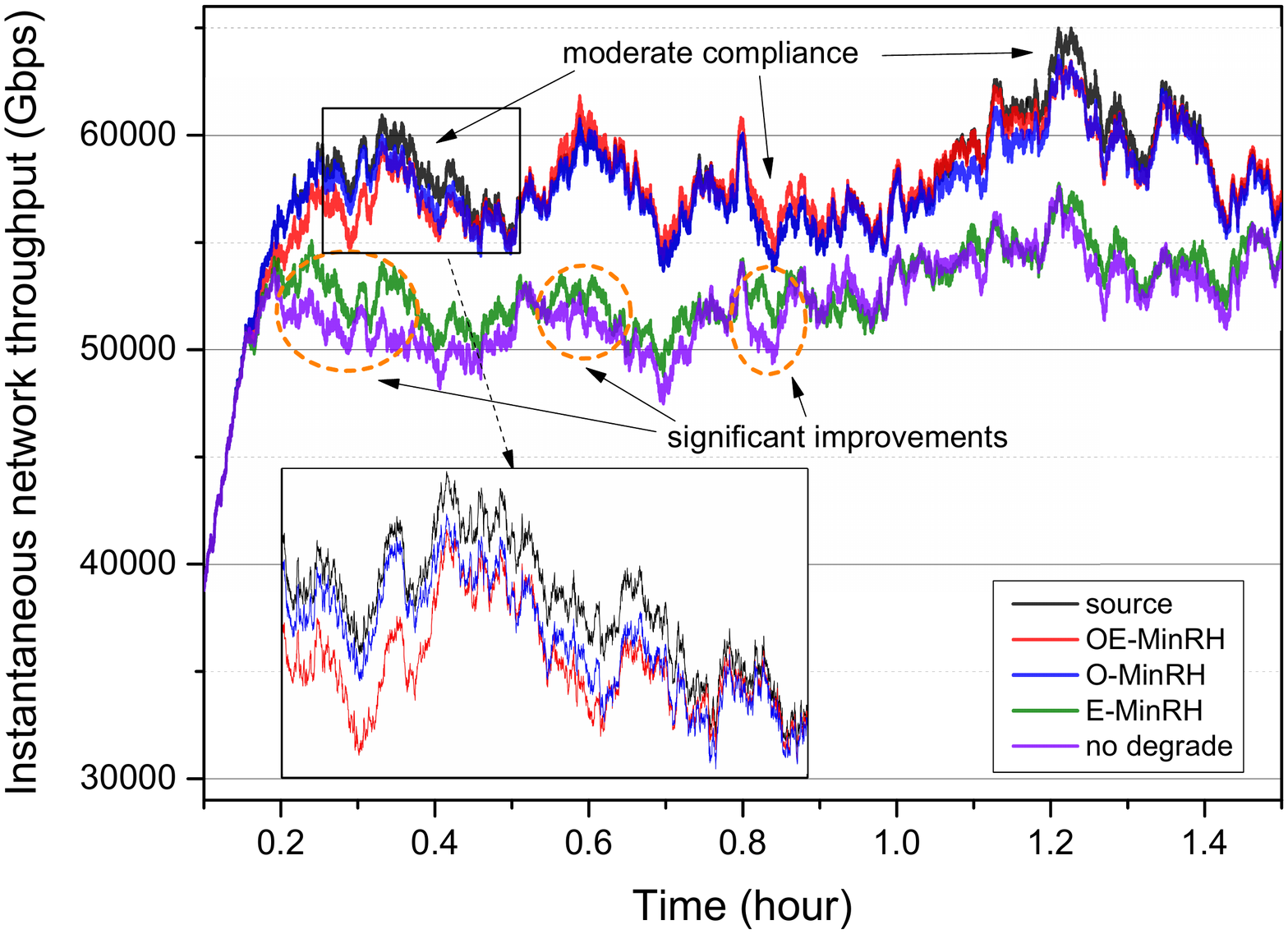}}
\hspace{0.15em}
\subfigure[Instantaneous BBP.]{
\label{fig:subfig:a} %% label for first subfigure
\includegraphics[width=2.2in]{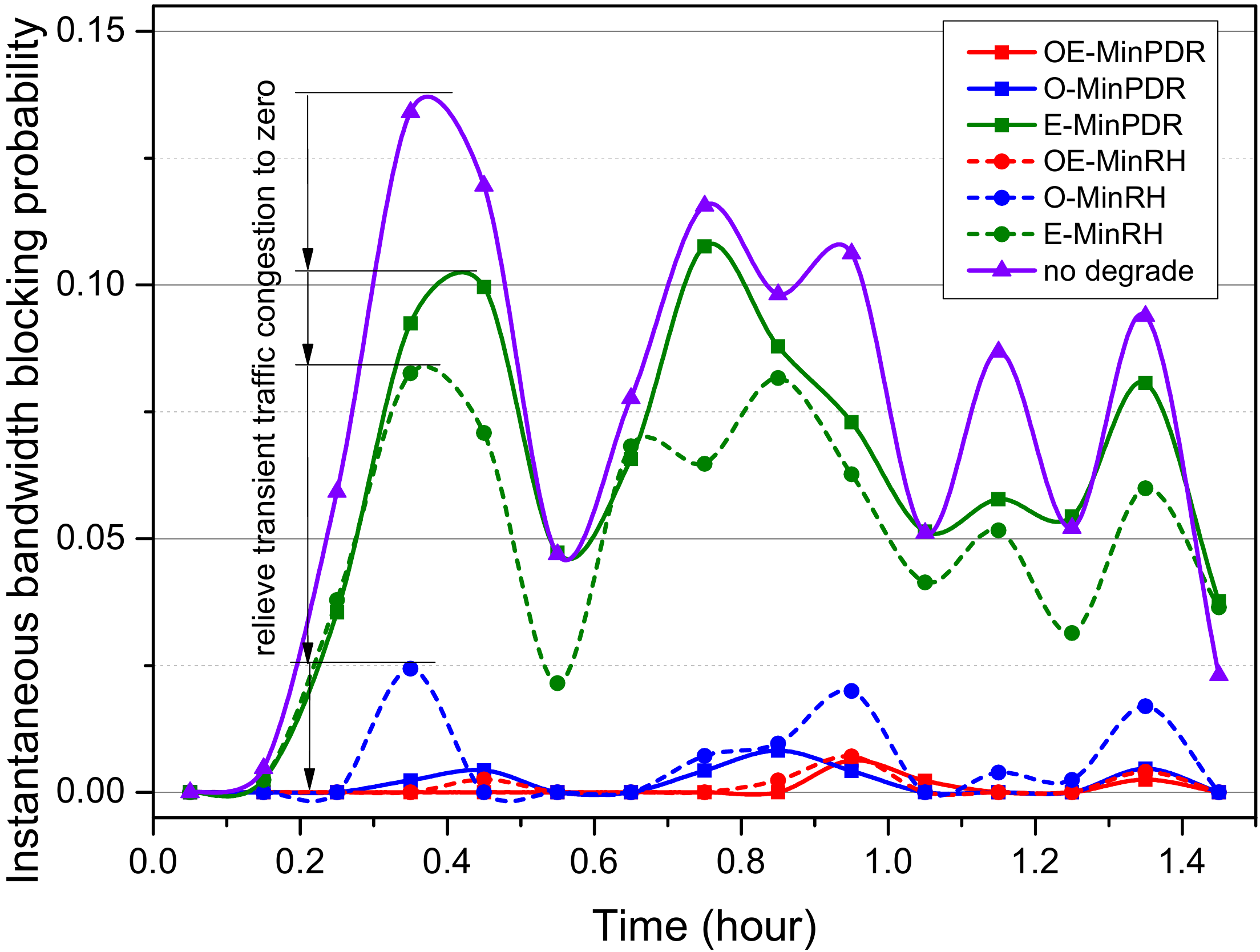}}
\vspace{-0.5em}
\hspace{-2em}
\caption{Transient analysis within 1.5 $hour$ (30 $Erlang$ per node, $\lambda=300\ hour^{-1}, \mu =10\ hour^{-1}$): different degradation policies.}
\label{fig:subfig} %% label for entire figure
\vspace {-1.5em} 
\end{figure*}

\begin{table}[htbp]
\vspace {-0.6em}
\renewcommand{\arraystretch}{1}
\vspace {-1em} 
\caption{Modulation format vs. Data rate vs. Transmission reach}
\label{table_example}
\centering
\vspace {-1em} 
\begin{tabular}{m{71pt}<{\centering} m{50pt}<{\centering} m{23pt}<{\centering} m{23pt}<{\centering} m{23pt}<{\centering}}
\hline 
Modulation format & BPSK~(default) & QPSK & 8QAM & 16QAM\\
\hline
Modulation~level & 2 & 4 & 8 & 16 \\

Bits~per~symbol & 1 & 2 & 3 & 4\\

Slot~bandwidth~(GHz)& 12.5 & 12.5 & 12.5 & 12.5\\

Data~rate~(Gbps)& 12.5 & 25 & 37.5 & 50\\

Transmission~reach~(km) & 9600 & 4800 & 2400 & 1200\\
\hline
\end{tabular}
\vspace {-1.5em}
\end{table}

\subsection{Dynamic Analysis}
Fig. 4 shows the bandwidth blocking probability (BBP) advantages of our proposed scheme over conventional scheme (threshold-based grooming \cite{wan2012}, no degradation). Fig. 4(a) shows the overall performance of all requests, and we can find there is a crossing point between optical degradation and both-layer degradation. In low-load area (26-34 $Erlang$), both-layer degradation (OE-MinPDR, OE-MinRH) performs the best, up to two orders of magnitude, while in high-load area (36-44 $Erlang$), optical-layer degradation (O-MinPDR, O-MinRH) performs the best. The reason is that, in high-load conditions, electric degradation (E-MinPDR, E-MinRH) achieves worse BBP than no degradation, which affects the blocking reduction by optical degradation in both-layer degradation. Figs. 4(b) and 4(c) show the BBP performance of requests in the highest and lowest priority. And we can conclude that all degradation policies can achieve significant blocking reduction for highest priority, while, for lowest priority, the blocking performance acts similar as requests with all priorities do. 

We also observe some common patterns in these three graphs. First, optical degradation performs almost the same regardless of priorities, because optical degradation does not involve service priorities as electric degradation does. Second, MinPDR performs better in optical-related degradations (both-layer degradation and optical degradation), while MinRH performs better only in electric degradation. This is because the route MinPDR returns tends to have a smaller number of existing requests, which increases the elements in ASSI or SBTL, thus increasing possibility of successful optical degradation. Also, the route that MinRH returns tends to have more existing requests, which increases the possibility of successful electric degradation. \emph{Actually, different mechanisms of optical degradation and electric degradation determine that MinPDR performs better for optical degradation, while MinRH suits electric degradation better}. The result that both-layer degradation and optical degradation performs similarly reveals that optical degradation has stronger influence on blocking reduction because it can enlarge the network capacity by high-order modulation, while electric degradation just deals with the bandwidth-time exchange to trade time for bandwidth.

\vspace {-0.5em} 
\subsection{Transient Analysis}
We conduct transient analysis on instantaneous network throughput and BBP. From Figs. 5(a) and 5(b), we obtain similar conclusions as the dynamic evaluations, that optical-related degradation achieves better compliance with the offered load in MinPDR, while electric degradation accomplishes better improvements in MinRH. Fig. 5(c) shows the instantaneous BBP variance over time, and we observe that different levels of blocking reduction can be achieved by different degradation policies. Both-layer degradation policies have the largest blocking reduction, and OE-MinPDR performs even better (almost zero blocking).
\thispagestyle{fancy}
\chead{\small Accepted, IEEE Globecom 2016, Optical Networks and Systems (ONS) Symposium, for ArXiv only.}
\vspace {-0.2em} 
\section{Conclusion}
In this work, we investigated dynamic QoS-assured degraded provisioning in service-differentiated multi-layer networks with optical elasticity. We proposed and leveraged the enhanced multi-layer architecture to design effective algorithms for network performance improvements. Numerical evaluations showed that we can achieve significant blocking reduction, up to two orders of magnitude. We also conclude that optical-related degradation achieves better with MinPDR, while electric degradation has lower blocking with MinRH due to different mechanisms of multi-layer degradations.


\IEEEtriggeratref{20}


\bibliographystyle{IEEEtran}

\vfill

\end{document}